 \documentclass[traditabstract,rnote]{aa}                                   

\usepackage{graphicx}
\usepackage{txfonts}
\begin{document}
   \title{Brans-Dicke model constrained from Big Bang nucleosynthesis and magnitude redshift relations of Supernovae} 


   \author{E. P. B. A. Thushari
          \inst{1},
          R. Nakamura\inst{1},
          M. Hashimoto\inst{1}, 
          and K. Arai\inst{2}, 
          }

   \institute{Department of Physics, Kyushu University,
	               Fukuoka, 812-8581, Japan\\
              \email{berni@phys.kyushu-u.ac.jp}\\
              \email{riou@phys.kyushu-u.ac.jp}\\
              \email{hashimoto@phys.kyushu-u.ac.jp}\\
            \and
             Department of Physics, Kumamoto University,
	               Kumamoto, 860-8555, Japan\\
             \email{arai@sci.kumamoto-u.ac.jp}
             }


 
  \abstract 
   {The Brans-Dicke model with a variable cosmological term ($BD\Lambda$) has been investigated with use of the coupling constant of $\omega=10^4$. Parameters inherent in this model are constrained from comparison between Big Bang nucleosynthesis and the observed abundances. Furthermore, the magnitude redshift ($m-z$) relations are studied  for $BD\Lambda$ with and without another constant cosmological term in a flat universe. Observational data of Type Ia Supernovae are used in the redshift range of $0.01<z<2$. It is found that our model with energy density of the constant cosmological term with the value of 0.7 can explain the SNIa observations, though the model parameters are insensitive to the $m-z$ relation.}



   \titlerunning{Brans-Dicke model constrained from $m-z$ relation}
   \authorrunning{E. P. B. A. Thushari et al.} 

   \maketitle
   

\section{Introduction}

Astronomical observations indicate that the cosmological constant in the very early universe exceeds the present value by some 120 orders of magnitude, which is estimated in modern theories of elementary particles (Weinberg \cite{Weinberg}). This is one of the fine tuning problems in cosmology called by the ``cosmological constant problem''. To explain the puzzle in cosmology, new modified theories  are needed beyond the standard model. That behavior of the cosmological term has motivated various functional forms to the cosmological term. The mechanism to the dynamical reduction of the cosmological term is formulated as a time dependent function (Silviera \& Waga \cite{Silviera}) and in terms of a scalar field (Weinberg \cite{Weinberg}, Huterer \& Turner \cite{Huterer}). On the other hand, generalized scalar tensor theories have been investigated (Wagoner \cite{Wagoner}, Endo \& Fukui \cite{Endo}, Fukui et al. \cite{Fukui}). 

Among them a Brans-Dicke (BD) theory with a variable cosmological term ($\Lambda$) as a function of scalar field ($\phi$) (Endo \& Fukui \cite{Endo}) has been proposed. This model has been investigated for the early universe and constrained from Big Bang nucleosynthesis (BBN) (Arai et al. \cite{Arai}, Etoh et al. \cite{Etoh}, Nakamura et al. \cite{Nakamura}) for the coupling constant $\omega\leq500$. Present observations suggest that the value of $\omega$ exceeds 40,000 (Berti et al. \cite{Berti}, Bertotti et al.  \cite{Bertotti}). Therefore, it is worthwhile to reconstrain the parameters in the Brans-Dicke model with  a variable cosmological term ($BD\Lambda$) for a new value of $\omega$. $BD\Lambda$ has played a very important role to explain the characteristics of the early universe (Arai et al. \cite{Arai}, Etoh et al. \cite{Etoh}, Nakamura et al. \cite{Nakamura}). However, an answer is needed to the question "How this model works at the present epoch?".
Therefore we adopt the magnitude redshift ~($m-z$) relations of Type Ia Supernova (SNIa) observations. This is because, the cosmological term affects the cosmic expansion rate of the universe significantly even at the low redshifts. SNIa observations imply that the universe is accelerating around the present epoch (Perlmutter et al. \cite{Perlmutter}, Riess et al. \cite{Riess}, \cite{Riess1}). 

In Sect. 2 formulation of $BD\Lambda$ is reviewed. Parameters inherent in this model are constrained in Sect. 3 from Big Bang nucleosynthesis for $\omega=10^{4}$. In Sect. 4 the $m-z$ relation is investigated for $BD\Lambda$ with including another constant cosmological term in a flat universe. Recent SNIa observational data (Astier et al. \cite{Astier}, Riess et al. \cite{Riess2}, Kessler et al. \cite{Kessler}) are adopted to constrain the models. Concluding remarks are given in Sect. 5.
\vspace{1mm}

\section{Brans-Dicke model with a variable cosmological term}
\vspace{0.5mm}
The field equations for $BD\Lambda$ are written as follows (Arai et al. \cite{Arai}): 
\begin{eqnarray} \label {p13_ee}\nonumber
R_{\mu\nu}-\frac{1}{2}g_{\mu\nu}R+g_{\mu\nu}\Lambda & = & \frac{8\pi}{\phi}T_{\mu\nu}+\frac{\omega}{\phi^{2}}\left(\phi_{,\mu;\nu}-\frac{1}{2}g_{\mu\nu}\phi_{,\alpha}\phi^{,\alpha}\right) \\
& & + \frac{1}{\phi}\left(\phi_{,\mu;\nu}-g_{\mu\nu}\Box\phi\right),
\end{eqnarray}
\vspace{0.2mm}
\begin{equation}\label{aa}
R-2\Lambda-2\phi\frac{\partial\Lambda}{\partial\phi}=\frac{\omega}{\phi^{2}}\phi_{,\nu}\phi^{,\nu}-\frac{2\omega}{\phi}\Box\phi,
\end{equation}where $\phi$ is the scalar field and $T_{\mu\nu}$ is the energy-momentum tensor of the matter field.
\vspace{0.5mm}
  The Robertson-Walker metric for homogeneous and isotropic universe is written as (Weinberg \cite{Weinberg1}):

\begin{equation}\label{aaa}
ds^{2}=-dt^{2}+a\left(t\right)^{2}\left[{\frac{{d}r^{2}}{1-kr^{2}}+r^{2}d\theta^{2}+r^{2}sin^{2}\theta d\phi^{2}}\right],
\end{equation}where $a(t)$ is the scale factor and $k$ is the curvature constant. Here we adopt $c=1$. 
The expansion is described by the following equation derived from the $(0,0)$ component of Eq. (\ref{p13_ee}):

\begin{equation}\label{p14_d}
  \left(\frac{\dot{a}}{a}\right)^{2}=\frac{8\pi\rho}{3\phi}-\frac{k}{a^{2}}+\frac{\Lambda}{3}+\frac{\omega}{6}\left(\frac{{\dot{\phi}}}{{\phi}}\right)^{2}-\frac{\dot{a}}{a}\frac{{\dot{\phi}}}{{\phi}},
   \end{equation}where $\rho$ is the energy density.\\
   
 We adopt the simplest case of the coupling between the scalar and matter field is
   
   \begin{equation}\label{dddd}
   \Box\phi=\frac{8\pi\mu}{2\omega+3}T^{\nu}_{\nu},
	\end{equation}where $\mu$ is a constant. Assuming a perfect fluid for $T_{\mu\nu}$, Eq. (\ref{dddd}) reduces to the following:
	
	\begin{equation}\label{eeee}
    \frac{d}{dt}\left(\dot{\phi}a^{3}\right)=\frac{8\pi\mu}{2\omega+3}\left(\rho-3p\right) a^{3},
    \end{equation} where $p$ is the pressure. 
	
A particular solution of Eq. (\ref{aa}) is obtained from Eqs. (\ref{p13_ee}) and (\ref{dddd}):  

\begin{equation}\label{p14_h}
 \Lambda=\frac{2\pi\left(\mu-1 \right)}{\phi}\rho_{m_{0}}a^{-3}, 
 \end{equation}where $\rho_{m_{0}}$ is the matter density at the present epoch.
 
  The gravitational "constant" $G$ is expressed as follows
\begin{equation}\label{p14_hh}
 G=\frac{1}{2}\left(3-\frac{2\omega+1}{2\omega+3}\mu\right)\frac{1}{\phi}.  
 \end{equation}     
  
  The density $\rho$ and the pressure $p$ are replaced by 
 \begin{eqnarray}\label{cccc}
\rho & = & \rho_{m}+\rho_{\gamma},\\
p & = & p_{\gamma}=\rho_{\gamma}/3
\end{eqnarray} where the energy density of matter varies as $\rho_{m}={\rho_{m}}_{0}a^{-3}$. The energy density of radiation is written as   $\rho_{\gamma}=\rho_{\gamma_{0}}a^{-4}$ except $e^{\pm}$ epoch:
$\rho_{\gamma}=\rho_{rad}+\rho_{\nu}+\rho_{e^{\pm}}$ at $t\leq1\rm{s}$, where subscripts $rad$, $\nu$ and $e^{\pm}$ are for photons, neutrinos and~ electron-positrons, respectively (Nakamura et al. \cite{Nakamura}). Subscript "0" indicates the values at the present epoch.\\

  Then, Eq. (\ref{eeee}) is integrated to give

 
   \begin{equation}\label{p14_g}
     \dot{\phi}=\frac{1}{a^{3}}\left[\frac{8\pi\mu}{2\omega +3}\rho_{{m}_{0}}t+B \right],   
\end{equation}where $B$ is an integral constant and here we use the normalized value of $B$: $B^{*}=B/(10^{-24} {\rm{ g \ s \ cm^{-3}}})$.

\vspace{3mm}

\begin{figure}
   \centering
  \rotatebox{-90}{\includegraphics[width=8.3cm,height=9cm]{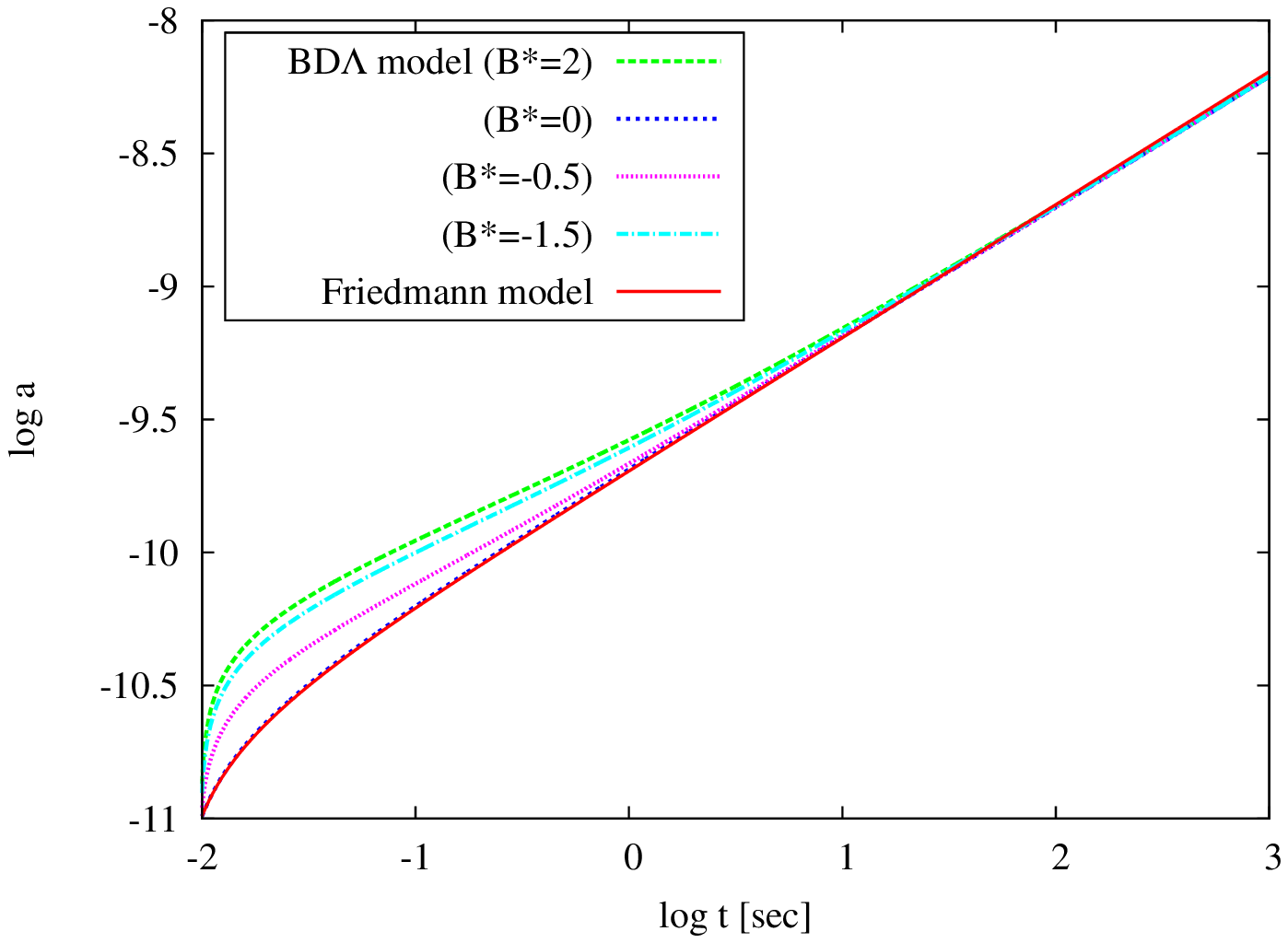}}
  \vspace{0.4cm}
     \caption{Time evolutions of the scale factor in $BD\Lambda$ with $\mu=0.6$ which are compared to the Friedmann model.
             }
        \label{a-t}
   \end{figure}
  Original Brans-Dicke theory is deduced for $\mu=1$ and it is reduced to the Friedmann model when $\phi$ = constant and $\omega\gg1$.
Physical parameters have been used to solve Eqs. (\ref{p14_d}), (\ref{p14_h}) and (\ref{p14_g}):  $G_{0} = 6.6726\times 10^{-8} {\rm{ cm^{3} g^{-1}s^{-2}}}$, $H_{0}=71 {\rm{ km \ s^{-1} \  Mpc^{-1}}}$ (Spergel et al. \cite{Spergel}), and $\omega=10^{4}$ (Berti et al. \cite{Berti}, Bertotti et al.  \cite{Bertotti}). Figure \ref{a-t} shows the evolution of the scale factor in $BD\Lambda$ for the several values of $B^{*}$. We identify considerable deviations in $BD\Lambda$ from the Friedmann model at $t<100$ s, which depends on the specific parameters. Therefore $BD\Lambda$ should be constrained from BBN (Arai et al. \cite{Arai}, Etoh et al.  \cite{Etoh}, Nakamura et al. \cite{Nakamura}).

\section{Parameters constrained from Big Bang nucleosynthesis}

\begin{figure}[htp]
\vspace{0.9cm}
\centering
\rotatebox{-360}{\includegraphics[width=8.7cm,height=9.6cm]{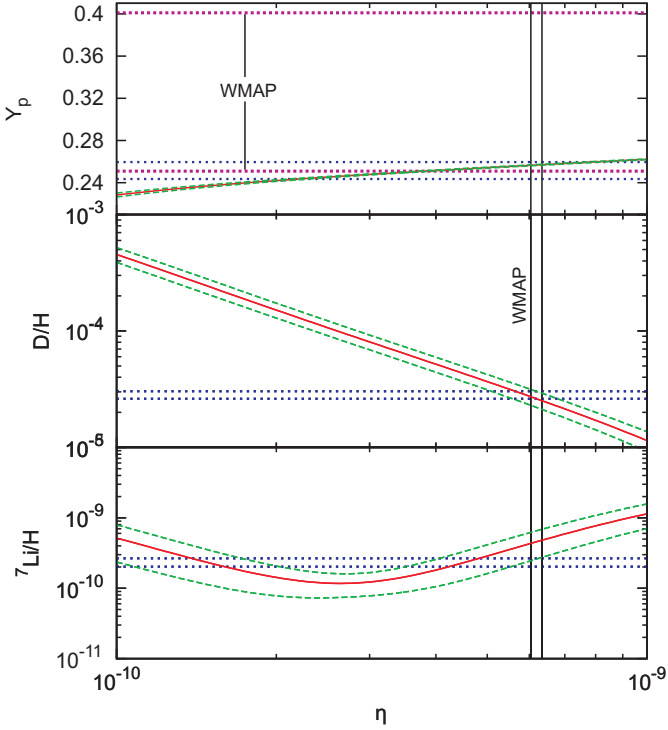}} 
\vspace{0.6cm}
\caption{Light element abundances of $\rm{{}^{4}He}$, $\rm{D}$, and $\rm{{}^{7}Li}$ vs. $\eta$ for $B^{*}=2$, $\mu=0.6$, and $\omega=10^{4}$. Dashed lines indicate the $\pm 2\sigma$  uncertainties in nuclear reaction rates in each abundance. The horizontal dotted lines indicate the regions of observational abundances. The solid vertical lines indicate the baryon-to photon ratio $\eta$.} 
\label{bbn}
\end{figure}
Big Bang nucleosyntheis provides powerful constraints on possible deviation from the standard cosmology (Malaney $\&$ Mathews. \cite{Malaney}). As shown in Fig. \ref{a-t}, the expansion rates of $BD\Lambda$ differs significantly from that of the standard Friedmann model. 

  The abundance of light elements in $BD\Lambda$ has already been investigated  (Arai et al. \cite{Arai}, Etoh et al.  \cite{Etoh}, Nakamura et al. \cite{Nakamura}).
In the previous studies, the parameters inherent in $BD\Lambda$ have been constrained for $\omega=500$. But we consider the case $\omega=10^{4}$ for convenience, because the Cassini measurements of the Shapiro time delay indicate $\omega\geq4\times10^{4}$ (Berti et al. \cite{Berti}, Bertotti et al.  \cite{Bertotti}). The detailed method of nucleosynthesis is described in Nakamura et al. \cite{Nakamura}.

  Figure \ref{bbn} shows the calculated abundances of $\rm{{}^{4}He}$, $\rm{D}$, and $\rm{{}^{7}Li}$ 
for $B^{*}=2$ and $\mu=0.6$.
The $\pm 2\sigma$ uncertainties in nuclear reaction rates
are indicated by the dashed lines. The horizontal dotted lines indicate the observational values of $\rm{{}^{4}He}$, $\rm{D/H}$, and $\rm{{}^{7}Li/H}$ as follows: ${Y_{p} = 0.2516 \pm 0.0080}$ (Fukugita \& Kawasaki \cite{Fukugita}), ${Y_{p}= 0.326\pm0.075}$ (Komatsu et al. \cite{Komatsu}), $\rm{D/H = (2.82\pm0.21)\times10^{-5}}$ (Pettini et al. \cite{Pettini}), $\rm{{}^{7}Li/H = (2.34\pm0.32)\times10^{-10}}$ (Melendez \& Ramirez. \cite{Melendez}). Here two observational values of $\rm{{}^{4}He}$ are used. The solid vertical lines indicates the WMAP constraint of the baryon-to-photon ratio,  $\rm{\eta=(6.19\pm0.15)\times10^{-10}}$ (Komatsu et al. \cite{Komatsu}).

  The intersection range of the two observational values of $\rm{{}^{4}He}$ is used to constrain the parameters.   
It is found that the values of $\eta$ derived from $\rm{{}^{4}He}$ and $\rm{D/H}$ are tightly consistent with 
the value by WMAP, though the lower limit of $\rm{{}^{7}Li/H}$ is barely consistent.
These agreements lead us to obtain the parameter ranges of $0.0\leq~\mu\leq0.6$ and $-2\leq B^{*}\leq2$.


\section{$m-z$ relation in $BD\Lambda$ with and without a constant cosmological term}

The distance modulus $\mu_{th}$ of the source at the redshift $z$ is

\begin{equation}\label{P14_m}
  \mu_{th}=m-M=5{\rm{log}_{10}}\left[\left(1+z\right)r_{l}\right]  +25,  
 \end{equation} 
where $m$ and $M$, are the apparent and absolute magnitudes, respectively and $r_{l}$ stands for the radial distance in units of $\rm{Mpc}$.

We adopt the SNIa (Astier et al. \cite{Astier}, Riess et al. \cite{Riess2}, Kessler et al. \cite{Kessler}) for which $\chi^{2}$ is defined by

\begin{equation}\label{ggg}
\chi^{2}=\sum_{i}\frac{\left(\mu_{th,i}-\mu_{obs,i}\right)^{2}}{\sigma^{2}_{i}}, 
\end{equation}where $\mu_{th,i}$ is given by Eq. (\ref{P14_m}), $\mu_{obs,i}$ and $\sigma_{i}$ are the observed values of distance modulus and their uncertinities.

For the homogeneous and isotropic universe, the relation between the radial distance and the redshift is derived from the Robertson-Walker metric as (Weinberg \cite{Weinberg2})

\[
\int_0^{z}\frac{d z}{H}  = 
\left\{
    \begin{array}{ll}
    k^{-1/2}\sin^{-1}\left(\sqrt {k}r_{l}\right) & \mbox{$k=+1$},\\   
    r_{l} & \mbox{$k=0$},\\ 
  \mid k\mid ^{-1/2}\sinh^{-1}\left(\sqrt {\mid k\mid } r_{l}\right) & \mbox{$k=-1$},
   \end{array}
\right. 
\] where $H=\dot{a}/a$ is the expansion rate written from Eq. (\ref{p14_d}) as

\begin{equation}\label{p14_o}
 H=\left[\frac{1}{4}\left(\frac{\dot{\phi}}{\phi}\right)^{2}-\left(1+z \right)^{2}k+\frac{\Lambda}{3}+\frac{\omega}{6}\left(\frac{{\dot{\phi}}}{{\phi}}\right)^{2}+\frac{8\pi}{3}\frac{\rho}{\phi}\right]^{\frac{1}{2}}-\frac{1}{2}\frac{\dot{\phi}}{\phi}.
   \end{equation}We conclude from the WMAP results that we live in a closely geometrically flat universe (Dunckley et al. \cite{Dunckley}). The present matter density $\rho_{m_{0}}$ is obtained from  Eq. (\ref{p14_d}) as 
 \begin{equation}\label{p14_p}
  H_{0}^{2}=\frac{1}{3}\left(\frac{8\pi\rho_{{m}_{0}}}{\phi_{0}}+\Lambda_{0}\right)+\frac{\omega}{6}\left(\frac{\dot{\phi}}{\phi}\right)^{2}_{0}-\left(\frac{\dot{\phi}}{\phi}H\right)_{0},
 \end{equation}
  
\begin{equation}\label{p14_ww}
\rho_{{m}_{0}}={4\rho_{c}^{BD\Lambda}}/\left({\mu+3}\right),\
\rho_{c}^{BD\Lambda}={3\phi_{0}H_{0}^{2}}/{8\pi},
  \end{equation}where $\rho_{c}^{BD\Lambda}$ is the critical density of $BD\Lambda$.  


\vspace{5mm}

Using the analogy with the Lema\^{i}tre model, Eq. (\ref{p14_p}) is transformed as

\begin{equation}\label{p14_ppp}
\Omega_{{m}_{0}}+\Omega_{{\Lambda}_{0}}+\Omega_{{\phi}_{0}}=1.
\end{equation}Here, energy density parameters are defined as

\begin{equation}\label{p14_ww}
\Omega_{{m}_{0}}=\frac{\rho_{{m}_{0}}}{{\rho_{c}^{BD\Lambda}}},\
\Omega_{{\Lambda}_{0}}=\frac{(\mu-1)\rho_{{m}_{0}}}{4\rho_{c}^{BD\Lambda}},
 \end{equation}
 
 \begin{equation}\label{p14_wws}
  \Omega_{{\phi}_{0}}=\frac{\omega}{6H_{0}^{2}}\left(\frac{\dot{\phi}}{\phi}\right)_{0}^{2}-\left(\frac{\dot{\phi}}{\phi }\frac{1}{H_{0}}\right)_{0}.
 \end{equation}

The value $\Omega_{{\phi}_{0}}$ is found to be very small as $7.01\times 10^{-5}$ for $\mu=0.6$. If  we consider the absolute value of  $\Omega_{{\phi}_{0}}$ in the parameter range $0.0\leq\mu\leq0.6$, its contribution to Eq. (\ref{p14_ppp}) is always less than $10^{-5}$. Therefore as long as we consider the present epoch, contribution from $\Omega_{{\phi}_{0}}$ can be neglected. 

\begin{figure}
   \centering
  \rotatebox{360}{\includegraphics[width=9cm,height=8cm]{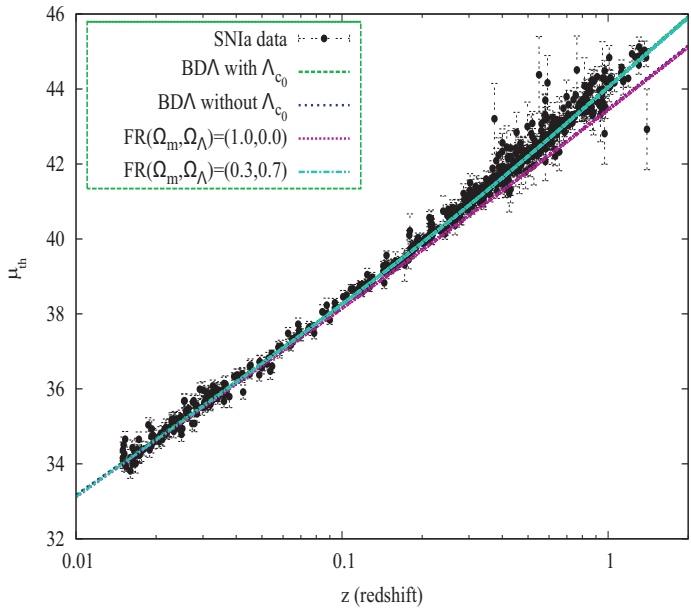}}
      \caption{Distance modulus vs. redshift for the flat universe in the Friedmann model and the $BD\Lambda$ with and without constant cosmological term constrained by SNIa observations (Riess et al. \cite{Riess2}, Astier et al. \cite{Astier}, Kessler et al. \cite{Kessler}).
              }
         \label{m-zall}
   \end{figure}

 \vspace{2mm}
Figure \ref{m-zall} shows the $m-z$ relation in $BD\Lambda$ for SNIa observations. Matter is dominant in this model. The energy density of the cosmological term is always less than $20\%$ in the best fit parameter region predicted in Sect. 3. The energy density of the cosmological term takes always negative values in the obtained parameter region. The parameter $B^{*}$ is not effective to change the values of $\Omega_{{m}_{0}}$ and $\Omega_{{\Lambda}_{0}}$. Since this model is matter dominant, it can not be constrained by the SNIa observations. \\

The Friedmann model with the energy density parameters of $(\Omega_{m},\Omega_{\Lambda})=(1.0,0.0)$, is merged with this $BD\Lambda$ model having reduced $\chi^{2}_{r}\equiv \chi^{2}/N\simeq 4.117$ (where $\chi^{2}=2293$ and $N$ is defined as degrees of freedom). This is inconsistent with the present accelerating universe, which should contain the sufficient amount of dark energy to accelerate the universe. To explain the present accelerating universe, it needs some modification to the cosmological term. 
\vspace{3mm}
  
  As the next approach, $BD\Lambda$ is modified by adding another constant cosmological term $\Lambda_{{c}_{0}}$. The expansion rate in this model is written by

\begin{equation}\label{p14_s}
 H=\left[\frac{1}{4}\left(\frac{\dot{\phi}}{\phi}\right)^{2}+\frac{\Lambda}{3}+\frac{\Lambda_{{c}_{0}}}{3}+\frac{\omega}{6}\left(\frac{{\dot{\phi}}}{{\phi}}\right)^{2}+\frac{8\pi}{3}\frac{\rho}{\phi}\right]^{\frac{1}{2}}-\frac{1}{2}\frac{\dot{\phi}}{\phi}.
\end{equation}

  The present matter density is

\begin{equation}\label{p14_ss}
 \rho_{{m}_{0}}=\frac{4\left(1-\Lambda_{{c}_{0}}\right)\rho_{{c}}^{{BD\Lambda}}}{\left(\mu+3 \right)}.
\end{equation}Here the energy density parameter of the constant cosmological term is fixed to be 0.7. \\
  
  We find that this model is consistent with the SNIa observations as seen in Fig. \ref{m-zall}. Total cosmological term becomes large in this model and consistent with the present accelerating universe with reduced $\chi^{2}_{r}\simeq 0.98$ (where $\chi^{2}=546.92$). For $\mu=0.5$, $BD\Lambda$ with $\Lambda_{{c}_{0}}$ predicts $\Omega_{\Lambda}=-4.285\times10^{-2}$ and $\Omega_{m}=0.343$. $\Omega_{\Lambda}$ always gets negative values in the parameter region of $\mu$ predicted in Sect. 3. If we consider the total value of energy densities, contribution from $\Omega_{\Lambda_{0}}+\Omega_{\Lambda_{{c}_{0}}}$ to the total energy density is always between $60\% - 67\%$. Therefore the cosmological term is dominant in the present epoch and it can be constrained from the present SNIa observations. It is concluded that  $BD\Lambda$ with $\Lambda_{{c}_{0}}$ has nearly the same energy density parameters as the Friedmann model with $(\Omega_{m},\Omega_{\Lambda})=(0.3,0.7)$. Although the cosmological term is not important at the early epoch, it plays very important role at the present era. All the parameters inherent in $BD\Lambda$ become insufficient as far as the $m-z$ relation is concerned.

\section{Concluding Remarks}

  Previous BBN calculations restricted the parameter range as $-0.5\leq\mu\leq0.8$ and $-10\leq B^{*}\leq10$ for $\omega=500$ (Nakamura et al. \cite{Nakamura}) . On the other hand, our large value of $\omega=10^{4}$ leads to decrease the parameter range of $B^{*}$ ($-2\leq B^{*}\leq-2$). It is oppositely affected the other parameter: $0.0\leq\mu\leq0.6$. The models parameters are inefficient in the $m-z$ relations of SNIa. \\

  In Sect. 4, the value of $\Omega_{{\phi}_{0}}$ is found to be much smaller compared with the other terms in Eq. (\ref{p14_ppp}).  Even though $\omega$ is increased until $10^{4}$ the contribution from $\Omega_{{\phi}_{0}}$ to Eq. (\ref{p14_ppp}) is always less than $1\%$ in the particular parameter range. There is no considerable wrong effect from the assumption we made in Sect. 4 to neglect the value of $\Omega_{{\phi}_{0}}$. In the parameter range $0.0\leq\mu\leq0.6$, $\Lambda$ has taken negative values according to Eq. (\ref{p14_ppp}). This may not conflict with theories, since the pressure of dark energy must be negative to reproduce the present accelerated expansion (Carroll \cite{Carroll}).\\ 
  
  It should be noted that from Eq. (\ref{p14_h}), $\Lambda \sim \rho_{m}/\phi$ and at the present epoch, $\Lambda_{0}$ is directly connected with $\rho_{{m}_{0}}$. Dark energy is written in terms of dark matter. However, dark energy and dark matter should be distinguishable to give rise to an accelerated expansion, since evolution of the scale factor seriously depends on the composition of each energy density of the universe. Therefore, $BD\Lambda$ without a constant cosmological term is indistinguishable from the matter dominant Friedmann model with the parameters of $(\Omega_{m},\Omega_{\Lambda})=(1.0,0.0)$.  It is noticed that the variable $\Lambda$ term in $BD\Lambda$ plays a minor role to accelerate the universe at the present epoch. Because of this reason, we have done a modification to $\Lambda$ by adding a constant cosmological term. It has no relation to  the expansion rate of the universe at the early epoch. However, the energy stored in the constant cosmological term has done a major role to accelerate the universe at the present epoch as seen in Fig. \ref{m-zall}. Since this model contains enough dark energy to accelerate the universe, it is constrained by the SNIa observations. In the present research, we have investigated $BD\Lambda$ at the early epoch to determine the intrinsic parameters and introduce new parameters at the present epoch for the $m-z$ relation. Since we have demonstrated a possibility of non-standard model which is compatible with the observations, it is worthwhile to examine more general functional form to the cosmological term (eg. Fukui et al.{\cite{Fukui}}).  


\end{document}